\newcommand{\tc}{\tilde{c}}

\documentstyle[aps,prl,twocolumn,epsfig,graphics,tabularx,array]{revtex}
\begin{document}

\twocolumn[\hsize\textwidth\columnwidth\hsize
\csname@twocolumnfalse\endcsname \draft
\title {Wannier functions analysis of the  nonlinear Schr\"{o}dinger equation with a periodic potential}

\author{G. L. Alfimov$^*$,  P. G. Kevrekidis$^{**}$, V. V. Konotop$^\dag$ and M. Salerno$^\ddag$ }
\address{
$^*$ F. V. Lukin's Institute of Physical Problems,
Zelenograd, Moscow, 103460, Russia.\\
$^{**}$ Department of Mathematics and Statistics, \\
University of Massachusetts, Amherst, MA 01003-4515, USA. \\
$^\dag$Departmento de F\'{\i}sica and Centro de F\'{\i}sica da
Mat\'eria
         Condensada, Universidade de Lisboa, \\
         Complexo Interdisciplinar, Av. Prof. Gama Pinto 2, Lisboa 1649-003,
         Portugal \\
         $^\ddag$Dipartimento di Fisica "E.R. Caianiello",
         Universit\'a di Salerno, I-84081 Baronissi (SA), Italy, and \\
         Istituto Nazionale di Fisica della Materia (INFM), Unit\'a
         di Salerno, Italy. }
\date{\today}

\maketitle

\date{\today}

\begin{abstract}
In the present  Letter we use the Wannier function basis to construct
lattice approximations of the nonlinear Schr\"{o}dinger
equation with a periodic potential. We show that the
nonlinear Schr\"{o}dinger equation with a
periodic potential is equivalent to
a vector lattice with long-range interactions. For the case-example of the
cosine potential we study the validity of the so-called  tight-binding
approximation i.e., the approximation when nearest neighbor
interactions are dominant. The results
are relevant to Bose-Einstein condensate theory as well as
to other physical systems  like, for example, electromagnetic
wave propagation in nonlinear photonic crystals.

PACS numbers: {42.65.-k, 42.50. Ar,42.81.Dp\\}

\end{abstract}

]

Interplay between nonlinearity and periodicity is the focus of numerous
recent studies in different branches of modern physics. The theory of
Bose-Einstein condensates (BEC) within the framework of the mean field
approximation \cite{BEC} is one of them. Recent interest in  the effects
of periodicity in BEC's has been stimulated by a series of remarkable
experiments realized with BEC's placed  in a potential created by a laser
field \cite{Kasevich} (the so-called {\it optical lattice}).
Nonlinearity and periodicity have been observed to introduce fundamental
changes in the properties of the system. On the one hand
periodicity modifies the spectrum of the underlying linear system
resulting in the potential of existence of new coherent
structures, which could not exist in a homogeneous nonlinear system.
On the other hand, nonlinearity renders  accumulation and transmission of
energy possible in ``linearly'' forbidden frequency domains; this, in
turn, results in field localization.
This situation is fairly general and can be found in other applications,
such as the theory of electromagnetic wave propagation in periodic media
(so-called photonic crystals) \cite{photonic}.

The study of nonlinear evolution equations with periodic
coefficients is a challenging and interdisciplinary problem. This
problem cannot be solved exactly in the general case and thus
gives rise to various approximate approaches. One of them,
borrowed from the theory of solid state \cite{solid}, is the
reduction of a continuous evolution problem to a lattice problem
(i.e., reduction of a partial differential equation to a
differential-difference one). It turns out that the relation
between the properties of periodic and discrete problems is indeed
rather deep (for a recent discussion of the relevant connections
see e.g., \cite{hetdis} and references therein). Following the
solid state terminology here we will refer to a discrete
approximation when only nearest neighbor interactions are taken
into account as a {\it tight-binding model}. This model has
recently been employed in the description of BEC in an optical
lattice~\cite{tromb}. One of the advantages of the lattice
approach is that it  allows one to obtain strongly localized
configurations, the so-called {\it intrinsic localized modes}
(ILM) (also called {\it breathers}) \cite{ILM}, in a rather simple
way. These entities correspond to {\it gap solitons} of the
original continuum model. In the above mentioned works a formal
analysis has been provided, using a basis of functions strongly
localized about the minima of the periodic potential. This basis,
however, has not been presented explicitly and even its existence
has not been established.

In this work  we propose to use Wannier-function (WF)
\cite{solid,kohn} as a complete set of functions localized near
the minima of the potential, to reduce the evolution of a
nonlinear partial differential equation with periodic coefficients
to a nonlinear lattice. WF has recently been used both in
connection with BEC in optical lattices \cite{tosi} and in
connection to gap solitons in nonlinear photonic crystals
\cite{konJOSA}. In our case this approach leads to {\it a vector
set} of lattice equations. These lattice equations {\it exactly}
correspond to the original continuum problem and the scalar
tight-binding approximation can be deduced from them under some
specific conditions. Checking these conditions one can analyse the
applicability of the tight-binding model. In particular, we argue
that although the ILM's reported in \cite{tromb} do exist, their
dynamics and stability must be studied within the framework of a
more general vector-lattice equation.

Being interested in BEC applications we base our analysis on the ubiquitous
example of the nonlinear Schr\"{o}dinger equation
\begin{equation}
\label{NLS}
 i\frac{\partial\psi}{\partial t}=-
 \frac{\partial^2\psi}{\partial x^2} +V(x)\psi+ \sigma |\psi|^2\psi
\end{equation}
where $\sigma=\pm 1$ and $V(x)$ is a periodic potential
$V(x+L)=V(x)$ \cite{remark}. Consider the eigenvalue problem
associated with (\ref{NLS})
\begin{equation}
-\frac{d^{2}\varphi _{k,\alpha }}{dx^{2}}+V(x)\varphi _{k,\alpha }=E_{\alpha
}(k)\varphi _{k,\alpha }  \label{eigen}
\end{equation}
where $\varphi _{k,\alpha }$ has  Bloch (Floquet) functions (BF's)
$\varphi _{k,\alpha }=e^{ikx}u_{k,\alpha }(x),$ with $u_{k,\alpha
}(x)$ periodic with period $L,$ and $\alpha $ is an index which
labels energy bands $E_{\alpha }(k)$. As is well known,
\cite{solid,kohn} $E_{\alpha }(k+\frac{2\pi }{L})=E_{\alpha }(k)$;
thus one can represent the energy as a  Fourier series
\begin{equation}
E_{\alpha }(k)=\sum_{n}\hat{\omega}_{n,\alpha \;}e^{iknL},\qquad
\hat{\omega}
_{n,\alpha }=\hat{\omega}_{-n,\alpha }=\hat{\omega}_{n\alpha }^{*}
\label{fourier}
\end{equation}
where an asterisk stands for complex conjugation and
\begin{equation}
\hat{\omega}_{n,\alpha }=\frac{L}{2\pi
}\int_{-\pi /L}^{\pi /L}E_{\alpha }(k)e^{-iknL}dk\,.
\end{equation}
The BF's constitute an orthogonal basis.
However, for our purposes it is more convenient to
use the WF's instead of the BF's. We recall that the
WF centered around the position $nL$ ($n$
is an integer) and corresponding to the band $\alpha$
is defined as
\begin{equation}
w_{\alpha }(x-nL)=\sqrt{\frac{L}{2\pi }}\int_{-\pi /L}^{\pi
/L}\varphi _{k,\alpha }(x)e^{-inkL}dk. \label{DefW}
\end{equation}
Conversely,
\begin{eqnarray}
\varphi _{k,\alpha }(x)=\sqrt{\frac{L}{2\pi }}\sum_{n=-\infty}^{\infty}w_{n,\alpha }(x)e^{inkL}.
\label{wannier1}
\end{eqnarray}
Similarly to BF's, they form a complete orthonormal (with respect
to both $n$ and $\alpha$) set of functions,
%\[
%\int_{-\infty}^{\infty} w_{n,\alpha }^{*}(x)w_{n^{^{\prime }},
%\alpha ^{^{\prime }}}(x)\,dx=\delta_{\alpha \alpha ^{^{\prime
%}}}\delta _{nn^{^{\prime }}}.
%\]
%
%\[
% \sum_{n,\alpha}w_{n,\alpha }^{*}(x^\prime)w_{n,\alpha}(x)=\delta(x-x^\prime).
%\]
%
which, by properly choosing the phase of the BF's in (\ref{DefW}),
can be made real and exponentially decaying at infinity
\cite{kohn}. In what follows we assume
 that this choice is made: $w_{n,\alpha }^{*}(x)=w_{n,\alpha}(x)$.
Due to completeness of WF's, any solution of (\ref{NLS}) can be
expressed in the form
\begin{equation}
\psi (x,t)=\sum_{n\alpha }c_{n,\alpha }(t)w_{n,\alpha }(x)  \label{expan1}
\end{equation}
which after substitution in (\ref{NLS}) gives
\begin{eqnarray}
i{\frac{dc_{n, \alpha}}{dt}}& =& \sum_{n_1}c_{n_1,
\alpha} \hat{\omega}_{n-n_1,\alpha } + \nonumber \\
&& \sigma \sum_{\alpha_1, \alpha_2, \alpha_3}\sum_{n_1,n_2,n_3}
  c_{n_1,\alpha_1}^{*} c_{n_2,\alpha_2} c_{n_3,\alpha_3} W^{n n_1 n_2
n_3}_{\alpha \alpha_1 \alpha_2 \alpha_3 }
\label{exact}
\end{eqnarray}
where
\begin{equation}
W^{n n_1 n_2 n_3}_{\alpha \alpha_1 \alpha_2
\alpha_3} = \int_{-\infty}^{\infty}
w_{n,\alpha}w_{n_1,\alpha_1}w_{n_2,\alpha_2}w_{n_3,\alpha_3} dx
\label{overlapp}
\end{equation}
are overlapping matrix elements.
Since WF's are real,
$W^{n_1 n_2n_3n_4}_{\alpha_1\alpha_2\alpha_3\alpha_4}$ is symmetric with
respect to all permutations within the groups of indices
$(\alpha,\alpha_1,\alpha_2,\alpha_3)$ and $(n,n_1,n_2,n_3)$.
Eq. (\ref{exact}) can be viewed as a vector discrete
nonlinear Schr\"{o}dinger (DNLS) equation for
${\bf c}_n=$col$(c_{n1},c_{n2},...)$ with long-range interactions.
In its general form, Eq. (\ref{exact})
is not solvable; however it allows reductions to simpler lattices in a
number of important special cases. Below we list some of them.

(i) If the coefficients of the Fourier series (\ref{fourier})
decay rapidly and $|\hat{\omega}_{1,\alpha}|\gg
|\hat{\omega}_{n,\alpha}|$, $n>1$ one can neglect  long-range
interaction terms in the linear part of Eq. (\ref{exact}) taking
into account nearest neighbors only.

(ii) Since $w_{n,\alpha}(x)$ is localized and centered around
$x=nL$, one can assume that in some cases among all the
coefficients $W^{n n_1 n_2 n_3}_{\alpha \alpha_1 \alpha_2
\alpha_3}$ those with $n=n_1=n_2=n_3$ are dominant and other terms
can be neglected. Then, taking into account  points (i), (ii) one
arrives at the equation
\begin{eqnarray}
i {\frac{dc_{n, \alpha}}{dt}} &=&\hat{\omega}_{0,\alpha}c_{n,\alpha}
+ \hat{\omega}_{1,\alpha }\left(c_{n-1,\alpha} +c_{n+1,\alpha}\right)+
\nonumber \\
&&+ \sigma \sum_{\alpha_1, \alpha_2, \alpha_3}
 W^{nnnn}_{\alpha \alpha_1 \alpha_2 \alpha_3 }  c_{n,\alpha_1}^{*}
 c_{n,\alpha_2} c_{n,\alpha_3}
\label{i-ii}
\end{eqnarray}
which degenerates into the tight-binding model \cite{tromb}
\begin{eqnarray}
i \frac{dc_{n,\alpha}}{dt} =\hat{\omega}_{0,\alpha}c_{n,\alpha}
+ \hat{\omega}_{1,\alpha}\left(c_{n-1,\alpha}
+c_{n+1,\alpha}\right)
\nonumber \\
+\sigma W^{nnnn}_{1111}  |c_{n,\alpha}|^2 c_{n,\alpha}
\label{TB}
\end{eqnarray}
if one restricts consideration to the band $\alpha$ only. Note
that within the single band approximation, Eq. (\ref{TB}) can be
generalized by including next nearest neighbor overlapping terms from Eq.
\ref{exact}, thus leading to the mixing of on-site and intra-site
nonlinearities of the same type as in the model introduced in
\cite{salerno92}. It should also be mentioned that the coefficients
$W^{nnnn}_{\alpha \alpha_1 \alpha_2 \alpha_3 }$ in Eq. (\ref{i-ii})
are {\it independent} of $n$.

(iii) In the general case, however, single band descriptions can
become inadequate (see below) due to resonant interband
interactions induced by nonlinearity (this is quite different from
{\it linear} solid state physics where interband transitions are
usually induced by external forces). In this case Eq. (\ref{i-ii})
can be further simplified by supposing that the periodic potential
depends on some parameter $\epsilon$: $V(x)\equiv V_\epsilon(x)$,
such that
$\hat{\omega}_{1,\alpha}\equiv\hat{\omega}_{1,\alpha}(\epsilon)
=O(\epsilon)$ when $\epsilon\to 0$.  After the transformation
$c_{n,\alpha}(t)=\exp\{i\hat{\omega}_{0,\alpha} t\}
\tilde{c}_{n,\alpha}(t)$ one arrives at the equation for
$\tilde{c}_{n,\alpha}$ with explicit dependence on $t$ in
the nonlinear terms in the form of oscillating exponents
$\exp[i(\hat{\omega}_{0,\alpha}+\hat{\omega}_{0,\alpha_1}-
\hat{\omega}_{0,\alpha_2}-\hat{\omega}_{0,\alpha_3})t]$. Let also
$\tilde{c}_{n,\alpha}(0)$ be small enough. Then on the timescale
$1/\epsilon$ these exponents are rapidly oscillating unless
$\alpha=\alpha_2, \alpha_1=\alpha_3$ or $\alpha=\alpha_3,
\alpha_1=\alpha_2$
%and time averaging
%can be used.
Then, denoting
$W^{nnnn}_{\alpha\alpha_1\alpha\alpha_1}\equiv  W_{\alpha\alpha_1}$
(the coefficients $W_{\alpha\alpha_1}$ do not depend on $n$ and describes
interband interactions), and
using time averaging techniques~\cite{proof},
one can reduce the lattice equation (\ref{i-ii}) to the form
\begin{eqnarray}
\nonumber
i\frac{d\tc_{n,\alpha}}{dt} =
\hat{\omega}_{1,\alpha}(\tc_{n-1,\alpha}&+&\tc_{n+1,\alpha})+\\
\label{lattice1} &+&\sigma\sum_{\alpha_1} W_{\alpha \alpha_1}
|\tc_{n,\alpha_1}|^2 \tc_{n,\alpha}.
\end{eqnarray}
This is a vector DNLS equation with coupling between bands of
the cross phase modulation type \cite{b3}. To investigate ILM
solutions in the Wannier representation we can restrict to the
scalar case described by Eq. (\ref{TB}) for which construction of
ILM's is well-established~\cite{ILM}. ILM's with multiple
components of $ \tc_{n,\alpha}$ populated can also be constructed
(see below).

Several comments about the above assumptions are in order. Firstly, the
latter imply
that the procedure of reduction of the NLS with periodic coefficients
to a lattice is a multistep process, and thus different lattices will
appear for different regions of the parameters. Secondly, for the reduction
to be consistent, the parameters of the problem must provide us with a
small parameter. Thus the largest of the quantities
$\hat{\omega}_{n,\alpha}/\hat{\omega}_{1,\alpha}$ ($n>1$) and
$W^{n n_1 n_2 n_3}_{\alpha \alpha_1 \alpha_2
\alpha_3}/W^{n n n n}_{\alpha \alpha_1 \alpha_2
\alpha_3}$ ($n_j\neq n$) will define this small parameter of the problem.
This, in particular, means that simplification of the lattice equation, and
hence the reasoning for the reduction to a lattice model, are 
(potentially) not always available 
for all parametric regimes, and must be verified for each model.

In the present Letter we study the validity of  the above
assumptions for Eq. (\ref{NLS}) with the potential $V(x)=A\cos (2x)$ 
(which corresponds
to the typical experimental setting for 
BEC in optical lattices~\cite{Kasevich}).
In this case Eq. (\ref{eigen}) is the Mathieu equation. Table I shows the
coefficients $\hat\omega_{n,\alpha}$  for the three lowest energy bands for
$A=-1$ and $A=-15$.

\bigskip

% \begin{table}
\begin{tabular}{|c||c|c|c||c|c|c|}
\hline & \multicolumn{3}{c||}{$A=-1$} & \multicolumn{3}{c|}{$A=-15$}
\\  \hline
 $n$ & $\alpha=1$  & $\alpha=2$ & $\alpha=3$   & $\alpha=1$  & $\alpha=2$
 & $\alpha=3$
\\ \hline \hline
$0$ & 0.1305 & 2.4657 & 6.3604   & -9.7862 & 0.0421 & 8.4659 \\
\hline
$1$ & -0.1428 & 0.5426 & -1.0067 & -0.0005 & 0.0151 & -0.1798 \\
\hline
$2$ & 0.0204 & 0.0784 & 0.0529 &    0.0000 & 0.0001 & 0.0091 \\
\hline
$3$ & -0.0048 & 0.0481 & -0.1107 & -0.0000 & 0.0000 & -0.0008 \\
\hline
$4$ & 0.0014 & 0.0225 & 0.0140 &    0.0000 & 0.0000 & 0.0001 \\
\hline
\end{tabular}

\vskip .2cm

{\small
TABLE I. The
first five Fourier coefficients $\hat\omega_{n\alpha}$ of the lowest
energy bands for two values of the amplitude potential.}

\bigskip

\noindent In general, one can conclude that the greater $|A|$ is,
the better the linear part obeys the nearest neighbour
approximation, which is intuitively expected since the probability
of tunelling between neighbor potential wells decreases with the
amplitude of the potential. At the same time, if $A$ is fixed the
coefficients  $\hat\omega_{n,\alpha}$, $n=0,\pm 1,..$ decay faster
for lower bands $\alpha$. The results illustrate, that the nearest
neighbor approximation works for both potential amplitudes,
while the averaging resulting in (\ref{lattice1}) is applicable
for $A=-15$ but not for $A=-1$. The reason is that in the latter case
the frequencies of oscillating exponents in (iii) are of order of
$\omega_{1,\alpha}$.

Moving to assumption (ii), let us introduce the following
notation. We denote by  $N_{\alpha,m}^{n}(\Delta)$ the number of
coefficients $W^{0 n_1 n_2 n_3}_{\alpha \alpha_1 \alpha_2
\alpha_3}$, $|n_i|\le n$, $\alpha_j\le m$, $i,j=1,2,3$ (the
coefficients with permuted indices are regarded as different) such
that $|W^{0 n_1 n_2 n_3}_{\alpha \alpha_1 \alpha_2
\alpha_3}|>\Delta$. As it is clear $\Delta$ plays the role of the
small parameter of the second condition, and
$N_{\alpha,m}^{n}(\Delta)$ gives the number of sites/zones
necessary to take into account for maintaining the given accuracy.
In the cases of the amplitudes $A=-1$ and $A=-15$ we have obtained
that $N_{1,1}^{n}(0.1)=N_{1,1}^{n}(0.01)=1$ for $n=1,...,5$. 
%Some
%other results for $N_{1,m}^{n}(0.1)$ and $N_{1,m}^{n}(0.01)$ are
%given in Table II.
For $N_{1,m}^{n}(0.1)$ and $N_{1,m}^{n}(0.01)$, see Table II.

\bigskip

\centerline{
\begin{tabular}{|c|c||c|c||c|c||}
\hline
$A$& $n$   & $N_{1,2}^n(0.1)$  & $N_{1,2}^n(0.01)$ &
$N_{1,3}^n(0.1)$  & $N_{1,3}^n(0.01)$ \\
\hline\hline
$-1$ &$0$  &  $4$ &  $4$ &  $7$ &  $13$ \\
\cline{2-6}
&$1$  &  $4$ &  $48$ & $7$ &  $219$ \\
\cline{2-6}
&$2$   &  $4$ &  $54$ & $7$ &  $249$ \\
\cline{2-6}
&$3,\,4$  &  $4$ &  $60$ & $7$ &  $303$ \\
\cline{2-6}
%&$4$  &  $4$ &  $60$ & $7$ &  $303$ \\
%\cline{2-6}
&$5$  &  $4$ &  $60$ & $7$ &  $339$ \\
\hline \hline
$-15$ &$0$  &  $4$ &   $4$ & $13$ &  $14$ \\
\cline{2-6}
&$1\,-\,5$ &  $4$ &  $4$ & $13$ &  $26$ \\
\hline \hline\end{tabular}
}
\bigskip
{\small TABLE II. The values $N_{\alpha,m}^{n}(\Delta)$.}

\smallskip

Table III
presents the overlapping coefficients $W_{\alpha\alpha_1}$ for two
values of the amplitude of the cos-like potential.

\bigskip

\centerline{
\begin{tabular}{|c||c|c|c|c|c|c|}
\hline
%& \multicolumn{6}{c|}{$\alpha\alpha_1$}\\
%\hline
A & $W_{11}$ & $W_{22}$  & $W_{33}$ & $W_{12}$ & $W_{13}$& $W_{23}$\\
\hline \hline
-1  & 0.375 & 0.240 & 0.173 & 0.182 & 0.152 &  0.142\\
\hline
-15  & 0.892 & 0.623 & 0.473& 0.417 & 0.262 & 0.326\\
\hline
\end{tabular}
}
\bigskip
{\small TABLE III. Overlapping coefficients $W_{\alpha\alpha_1}$.}

\bigskip

\begin{figure}[tbp]
\epsfxsize=7cm \epsffile{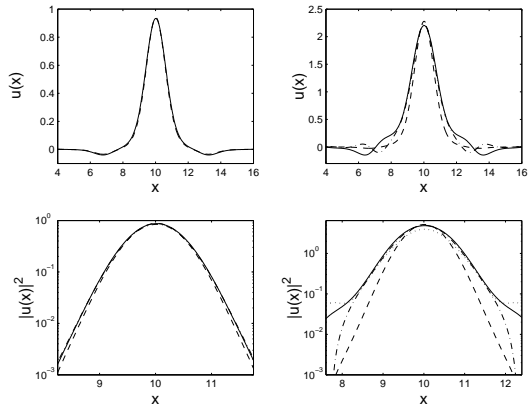}
\caption{Comparison of the
lattice reconstructed solution in the tight binding (dashed line)
and the 3-band (dash-dotted line) approximation with the exact
solution (solid line). The comparison is performed for $A=-5$ and
$\omega=-1.5$ (left panels) and $A=-5$, $\omega=1.5$ (right
panels). In the bottom (semilog) panels additionally the result of
dynamical time evolution of the tight binding approximation is
shown by the dotted line.
%Notice that for the
%bottom left panel the dotted line is practically coincident with the
%solid line.
The latter can be seen
to approach, as time evolves (the shown snapshots are for
$t \approx 50$), the shape of the exact solution (in the left panel
it can actually not be distinguished from it) and to match its
asymptotics, possibly shedding small wakes of low amplitude wave
radiation in the process (see e.g., the bottom right panel).}
\label{wfig1}
\end{figure}

\noindent It follows from Tables II and III that, in general {\it
one cannot neglect the contribution of the WF of the highest
zones}.
%which must be taken into account to improve the
%tight-binding approximation (\ref{TB}).
However, one can show that the model (\ref{TB}) can be
successfully used to describe bright monochromatic GS solutions of
(\ref{NLS}) of the form $\psi(t,x)=e^{i\omega t} u(x)$. An example
is shown in Fig. \ref{wfig1} for (the ``intermediate'' between the
above presented ones case of) $A=-5$. The two panels show the
cases of $\omega=-1.5$ (left panels) and $\omega=1.5$ (right
panels), for $\sigma=1$. The top panels show the comparison of the
exact solution (shown by solid line) of Eq. (\ref{NLS}) with the
reconstructed profile obtained from solving Eq. (\ref{lattice1})
and using Eq. (\ref{expan1}). The relevant profiles in the tight
binding approximation are shown by dashed line, while in the right
panel (where the one band approximation is less accurate), the
3-band approximation is also shown by dash-dotted line. The bottom
panels show in a semilog plot the square modulus of the
configurations of the top panels as well as, additionally, by
dotted line the result of time evolution (for $t \approx 50$) of
Eq. (\ref{NLS}) with the tight binding approximation as the
initial condition of the simulation. One can straightforwardly
observe that the approximate solution ``reshapes'' itself into the
exact solution (possibly shedding some very small amplitude
radiation wakes in the process). This demonstrates that the method
can be used very efficiently to construct (approximate) solutions
of the original PDE, by using the lattice reduction in the WF
representation.

Let us return now to the requirement (iii) and argue that choosing the
small parameter as $\epsilon=|A|^{-1}$ one can provide averaging
of (\ref{NLS})  in the limit $A\to-\infty$. Namely, we claim that

(a) If $\alpha$ is fixed and $A\to -\infty$
then
\begin{eqnarray*}
\hat{\omega}_{0,\alpha}\sim A+(2\alpha-1)\sqrt{-A}
+((2\alpha-1)^2+1)/8\,.
%\label{AS1}
\end{eqnarray*}
If $A$ is fixed then $\hat{\omega}_{0,\alpha}$ tends to
infinity as $\alpha$ grows;

(b) If $\alpha$ is fixed then the value
$\hat{\omega}_{1,\alpha}$ tends to zero faster than
any power of $1/|A|$; at the same time if $A$ is fixed
then $\hat{\omega}_{1,\alpha}$ tends to infinity as $\alpha$
grows;

(c) If $\alpha$ is fixed and $A\to -\infty$
the Wannier functions can be approximated by the formula
\begin{eqnarray*}
w_{0,\alpha}(x)\approx \frac{(2|A|)^{\frac{1}{8}}}
{\pi^{\frac{1}{4}}\sqrt{2^{\alpha-1}(\alpha-1) !}}
e^{-\sqrt{\frac{|A|}{2}} x^2} H_{\alpha-1}
\left((2|A|)^{\frac{1}{4}} x\right)
%\label{Her}
\end{eqnarray*}
where $H_k(y)$ are Hermite polynomials. This is a natural consequence
of the fact
that for  sufficently low levels the potential can be well approximated
by the parabolic one;

(d)  The coefficients $W_{\alpha\alpha_1\alpha_2\alpha_3}^{n n_1 n_2 n_3}$
with different $n$,$n_1$,$n_2$ and $n_3$ tend to zero as $A\to -\infty$ and
at the same time
%\begin{equation}
$
W_{\alpha\alpha_1\alpha_2\alpha_3}^{nnnn}\approx
K_{\alpha\alpha_1\alpha_2\alpha_3} |A|^{\frac{1}{4}}
$
%\label{W}
%\end{equation}
where $K_{\alpha,\alpha_1,\alpha_2,\alpha_3}$ do not depend on $A$
and can be expressed explicitly through the integrals of products
of Hermite polinomials \cite{slater}.

Taking into account (a)-(d), making the substitution
$c_{n,\alpha}(t)=e^{i\omega_{0,\alpha} t}
|A|^{-1/8} \tilde{c}_{n,\alpha}(t)
$
and averaging over rapid oscillations one arrives at (\ref{lattice1})
with $W_{\alpha\alpha_1}=K_{\alpha\alpha_1\alpha\alpha_1}$.

To conclude we have shown how to  derive lattice
models which approximate efficiently nonlinear partial differential
equations with periodic coefficients.
This analysis gives the possibility to control the validity
of the tight-binding approximation. In particular, we have shown that in a 
large region of parameter space, for the cos-like potential, one cannot 
restrict 
consideration to the lowest band. 
%(i.e., to the tight-binding approximation). 
This is due to interband transitions originating from the nonlinearity (a 
situation very different from the one known in the (linear) solid-state 
physics, where the interband transitions occur due to effect of 
perturbations). However, there exist parameter ranges where with reasonably 
high accuracy the atomic wave function (that is a bright gap soliton of the 
1D NLS equation) is approximated by a single WF. Such a state will form a 
``Wannier-soliton'' that should also be experimentally observable.
It should be highlighted that the use of the WF basis allows one to
test, extend and improve the tight-binding 
approximation, in a {\it controllable} and {\it systematic} fashion
by accounting for higher order terms in the Wannier expansion.    
Moreover, there is a computational gain when computing with a 
discrete system with respect to the corresponding cost for a much finer 
mesh (needed to resolve the original continuous system). While 
this gain may not be overly significant in one dimension,
it may prove quite useful in tackling higher dimensional problems.

It should be stressed that even though developed for a specific,
physically relevant (to optical lattices in BEC) setting, the
approach presented here is {\it very general} and
{\it directly  applicable}
to numerous other physical problems including  the description
of solitary wave  propagation
through one-dimensional photonic crystals,
\cite{AlKon}
chemical reactions on periodic catalytic
substrates,
\cite{yannis}
or even population dynamics
in appropriately heterogeneous substrates.
\cite{Teramoto}.

%%%%%%%%%%%%%%%%%%%%%%%%%%%%%%%%%%%%%%%%%%%%%%%%%%%%%%%%%%%%%%%

G.L.A thanks CFMC of the University of Lisbon, University of
Salerno for warm hospitality. 
The work of GLA has been partially
supported by the Senior NATO fellowship. VVK acknowledges support
from the European grant, COSYC n.o HPRN-CT-2000-00158. PGK
gratefully acknowledges support from a UMass Faculty Research Grant
and from the Clay
Foundation.
% through a Special Project Prize Fellowship. 
MS
acknowledges the MURST-PRIN-2000 Initiative and the European grant
LOCNET n.o HPRN-CT-1999-00163 for partial financial support.


\begin{references}
\bibitem{proof} A rigorous proof of the averaging procedure will be published elsewhere.
\bibitem{remark} For simplicity we restrict to the one dimensional case but the
analysis can be easily generalized to separable periodic
potentials in higher dimensions.
\end{references}
\end{document}